\documentclass[%
 reprint,
 onecolumn,
superscriptaddress,
nofootinbib,
 amsmath,amssymb,
 aps,
prd,
]{revtex4-2}

\usepackage{graphicx}
\DeclareGraphicsExtensions{.png,.pdf}
\usepackage{dcolumn}
\usepackage{bm}

\usepackage[dvipsnames]{xcolor}
\usepackage{hyperref}
\usepackage[utf8]{inputenc}
\usepackage{lineno}

\begin{document}

\title[cosmic ray deuteron]{Current status and new perspectives on cosmic ray deuterons}

\author{Diego Mauricio Gomez-Coral}
\email{dgomezco@hawaii.edu}
\email{dgomezco@fisica.unam.mx}
\affiliation{Department of Physics and Astronomy, University of Hawaii at Manoa, 2505 Correa Road, Honolulu, Hawaii 96822, USA} %
\affiliation{Instituto de Física, Universidad Nacional Autónoma de México Circuito de la Investigación Científica, Ciudad de México, México}
\author{Cory Gerrity}
\email{cgerrity@hawaii.edu}
\affiliation{Department of Physics and Astronomy, University of Hawaii at Manoa,
2505 Correa Road, Honolulu, Hawaii 96822, USA} %
\author{Riccardo Munini}
\affiliation{INFN, Sezione di Trieste,
Padriciano 99, 34149 Trieste, Italy} %
\author{Philip von Doetinchem}%
\affiliation{Department of Physics and Astronomy, University of Hawaii at Manoa,
2505 Correa Road, Honolulu, Hawaii 96822, USA} %

\date{\today}%


\begin{abstract}

Deuterons are the most abundant secondary cosmic ray species in the Galaxy, but their study has been severely limited due to experimental challenges. In an era with new experiments and high-precision measurements in cosmic rays, having a low-uncertainty deuteron flux in a wide energy range becomes possible. The deuteron-over-helium ratio ($d$/$^4$He) is important to understand the propagation of cosmic rays in the Galaxy and in the heliosphere, complementing observations with heavier nuclei like the boron-to-carbon ratio. In this work, the most up-to-date results of the deuteron flux and the $d$/$^4$He ratio at the top of the atmosphere have been obtained using {\tt GALPROP} and a 3D solar modulation model. It was found that the simulation describes the deuteron flux and $d$/$^4$He data below 1\,GeV/$n$ within the uncertainties of the model. However, the model underestimates the best-published measurements available at high energy. This discrepancy suggests different effective diffusions have to be considered between secondary light species like deuterons and heavier nuclei. Either this is a consequence of a break in the universality of propagation between light and heavier nuclei or a lack of precision measurements, it is something AMS-02 will help to resolve in the near future.

\end{abstract}

\maketitle

\newpage
\section{Introduction}\label{s1}

Hydrogen (H) and helium (He) isotopes are the most abundant species in cosmic ray (CR) nuclei. Protons ($p$) represent nearly 90\% of these charged particles, and it is estimated that CR deuterons ($d$), the only other stable H isotope, are approximately 2\%-3\% of the proton abundance\,\cite{BESS93, PAMELA}. Since CR deuterons are known to be destroyed in the nuclear processes that occur during stellar formation, they are not expected to be accelerated in supernova remnants like primary CRs such as protons, helium--4 ($^{4}$He), carbon (C), and oxygen (O) do. Instead, they are predicted to originate from fragmentation interactions between $^{4}$He, carbon (C), oxygen (O), and other heavier nuclei with the interstellar medium (ISM), i.e., they are considered secondary CRs\,\cite{Sina}. Because of their identical charge yet lower abundance compared to protons, deuterons are challenging to detect in cosmic ray experiments. However, the information obtained from their measurement has important implications for our understanding of the cosmic ray field.

Secondary CRs such as d, helium--3 ($^{3}$He), lithium (Li), beryllium (Be), boron (B), etc., are probes to characterize the propagation process of CRs in the Galaxy. Secondary-to-primary CRs ratios are directly related to the amount of material traversed by CRs. This quantity is known as grammage ($X$), and is usually reported to have a value of around 10\,$\text{\,g\,cm}^{-2}$ for particles with an energy of about 10\,GeV per nucleon\,\cite{Blasi2013,crreview}. This value is typically obtained from the more abundant boron-to-carbon ratio (B/C) measurements, given the experimental advantage of separating both elements by measuring the charge rather than separating H or He isotopes by their masses. The measured B/C ratios demonstrate the diffusive motion of CRs in the Galaxy, and are a key component in constraining the diffusion parameters. In the standard propagation scenario all CR species are driven by the same diffusion parameters obtained with the B/C ratio, however, the lack of information about other secondary-to-primary ratios, especially lighter nuclei like $d$/$^{4}$He and $^{3}$He/$^{4}$He, has not allowed rigorous testing of this ``universality in propagation" postulate. 

Fortunately, in the last years new experiments with better capabilities, like PAMELA\,\cite{PAMELA} and AMS-02\,\cite{AMS022015p}, are able to separate isotopes in a wide energy range. These experiments have been measuring other secondary-to-primary ratios and individual nuclei spectra successfully. This has revived a debate about a violation of the universality in propagation\,\cite{WEBBER1997, Tomassetti_2012, Coste, Korsmeier2020, Korsmeier2021}, exploring new scenarios like a nonhomogenous propagation medium\,\cite{J_hannesson_2016}. In this context, precision measurements of $d$/$^{4}$He or $d$/He in the region above 1\,GeV, where solar effects are less critical, will help the cosmic ray community to probe the universality of cosmic ray propagation through the Galaxy and clarify if there is any difference in the transportation process between light and heavy CR nuclei. On the other hand, more AMS-02 data at energies below 1\,GeV will shed light on deuteron CR propagation in the heliosphere, specifically by determining if there are differences in the diffusion processes followed by species with a similar mass number to charge ratio\,($A/Z$). This line of reasoning stems from the following: the AMS-02 measurements show a time dependence in the proton-to-helium ratio in the 1.92--2.15\,GV rigidity bin that is not present in higher rigidity bins. By varying the local interstellar spectrum (LIS) while treating $A/Z$ as a parameter set to be the same between protons, helium-3, and helium-4, the authors of \cite{Corti} conclude that this time dependence cannot be explained by a different LIS around 2\,GV. A possible resolution to this is based on the fact that particles with a different $A/Z$ have different velocities in the same rigidity bin, as the diffusion coefficients have a velocity dependence. Since the diffusion coefficients depend on rigidity and velocity, particles with a different $A/Z$ feel different diffusion effects. Thus, time dependence in the proton-to-helium ratio may be due to diffusion changing over time. This can be tested by studying particle ratios of the same $A/Z$, such as the deuteron-to-helium ratio, where a time dependence could not be attributed to a difference in $A/Z$. This is a strong motivating factor for precision deuteron-to-helium ratio measurements.  Furthermore, low-energy measurements with balloon-borne experiments like GAPS\,\cite{GAPS}, where spectrometers such as PAMELA and AMS-02 have no access, will help to improve solar modulation models and the understanding of secondary light nuclei produced by nuclear collisions in the upper atmosphere.

There are few but important recent studies in deuterons and light nuclei CRs\,\cite{Tomassetti_2012, Coste, Tomassetti_2017, WU2019292, Weinrich}, where main components such as the production cross sections, propagation parameters, as well as solar modulation effects, were reviewed and updated. The goal of this work is to explicitly show the impact of deuteron and deuteron-to-helium ratio measurements on interpreting the cosmic ray theory with associated uncertainties, using the most up-to-date high-precision data. In Sec.\,\ref{sec2}, a summary of the latest and most relevant measurements in CR deuterons for this study is presented. In Sec.\,\ref{sec3}, the dynamics of the current understanding of deuterons in cosmic ray propagation in the Galaxy and their modulation by Solar influences are described, emphasizing the role that this light nucleus plays in improving our understanding of these processes. Sec.\,\ref{sec4} discusses how deuterons and other light isotopes are produced in cosmic rays, the available measurements in accelerator experiments, and the most recent parametrization of cross sections for these particular reactions. In Sec.\,\ref{sec5}, the simulation results of the cosmic ray deuteron flux and deuteron-over-helium ratio are presented. These are based on comparisons to the latest high-precision cosmic ray measurements, as well as new cross section parametrizations for cosmic ray interactions. In Sec.\,\ref{sec6} final remarks and conclusions are given.

\section{CR deuteron measurements}\label{sec2}

\begin{figure*}[t]
\begin{center}
\begin{tabular}{ll}
\includegraphics[width=0.49\linewidth]{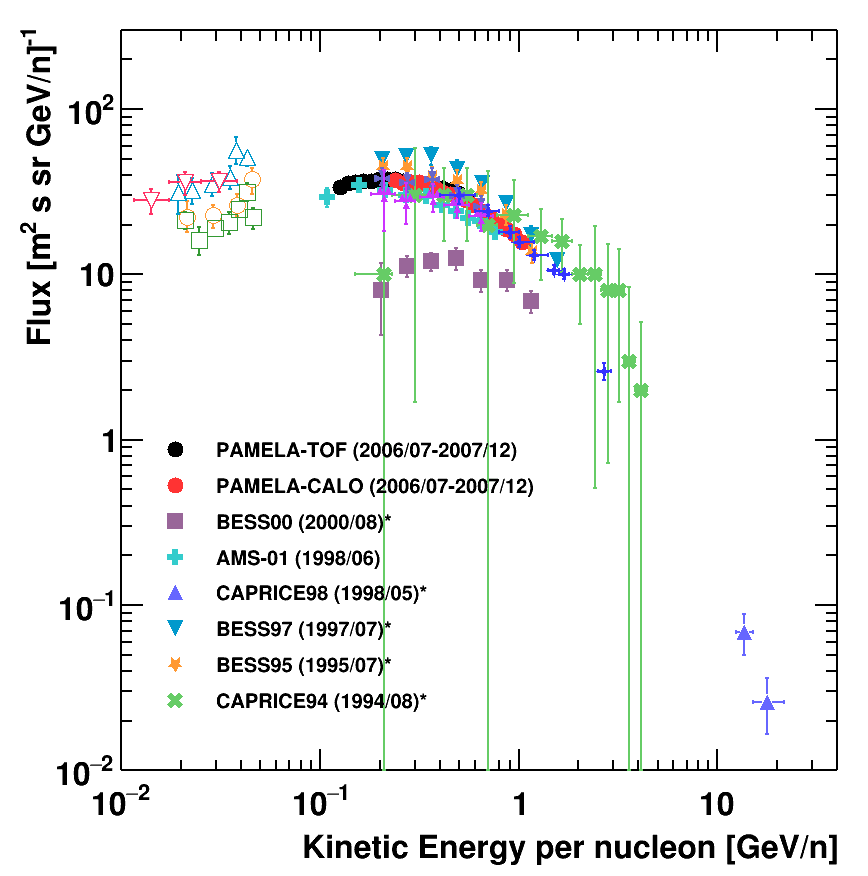}
& \includegraphics[width=0.49\linewidth]{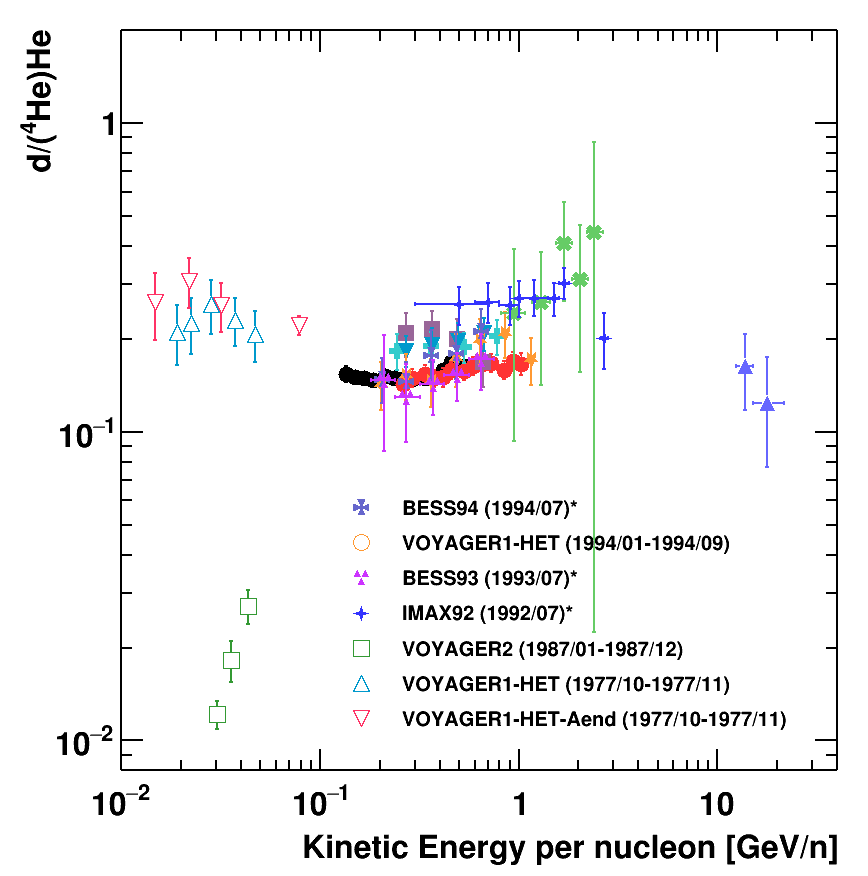}
\end{tabular}
\caption{Current status of deuteron measurements: Left plot shows deuteron flux measurements and right plot shows $d$/$^{4}$He and $d$/He(*) ratios taken in recent years. Figure has been adapted from\,\cite{crdb, crdbv21}.}\label{sec2:fig1}
\end{center}
\end{figure*}

Summaries of deuteron and deuteron-to-helium ratio measurements from seven of the most up-to-date cosmic ray experiments are presented in Fig.\,\ref{sec2:fig1}\,\cite{crdb, crdbv21}. Four of the results are the satellite-borne detectors PAMELA\,\cite{PAMELA}, AMS-01\,\cite{AMS01}, VOYAGER\,1\,\cite{Voyager1_77, Voyager1_94}, and VOYAGER\,2\,\cite{Voyager2}, and three of them are the balloon-borne detectors IMAX\,\cite{IMAX}, BESS\,\cite{BESS93, BESS939495, BESS97, BESS98, BESS00}, and CAPRICE\,\cite{CAPRICE94, CAPRICE}. These experiments measure CR deuterons using two different detection techniques. Magnetic spectrometers like PAMELA, AMS-01, BESS, and CAPRICE identify H isotopes by their mass ($m=RZe/\gamma \beta c$) using a direct measurement of the rigidity ($R=pc/Ze$), charge ($Ze$) and velocity ($\beta$). On the other hand, the interstellar probes VOYAGER\,1 and VOYAGER\,2 identify H and He isotopes by measuring their energy loss per unit path length (d$E$/d$x$) and their total kinetic energy.

From Fig.\,\ref{sec2:fig1} it is evident that deuteron measurements have, in general, high uncertainties at all energies ($>$20\%), with the only exceptions being the results from PAMELA with errors on the order of 10\%-15\%, and AMS-01 with flux errors below 10\% and deuteron-to-helium ratio errors on the order of 10\%-25\%. Additionally, the energy region above 1\,GeV/$n$ has not been explored entirely because of the inherent difficulty of separating H isotopes at high energies. For the case of $d$/$^{4}$He, measurements are less abundant than $d$/He considering again the challenge to separate He isotopes. In both cases, data have high uncertainties, especially above 1\,GeV/$n$. Figure\,\ref{sec2:fig1} also shows that most measurements are below 1\,GeV/$n$ where solar effects are important. Also, data were taken in reduced time periods, not allowing for a detailed analysis of the deuteron flux or $d$/$^{4}$He(He) ratios as a function of time. The combination of recent precision measurements by the PAMELA experiment, upcoming high precision results by the AMS-02 experiment\,\cite{AMS022015p}, and future measurements by balloon experiments like GAPS\,\cite{GAPS}, will fill some of the voids presented in Fig.\,\ref{sec2:fig1}, reduce the uncertainties, and measure for longer periods providing the possibility for time-dependent analysis.

An important source of uncertainty for experiments flown as a balloon payload is the interaction of CRs with the atmosphere above them. For a typical cosmic ray balloon experiment, the Earth's atmosphere provides a residual grammage on the order of $\text{5 g/cm}^2$\,\cite{BESS93}. This layer of propagation material is comparable to that experienced by a typical cosmic ray during its propagation through the Galaxy\,\cite{grammage}. Therefore, cosmic ray balloon measurements require the appropriate treatment of atmospheric effects to produce meaningful top-of-the-atmosphere (TOA) spectra. A typical atmospheric treatment for a balloon experiment (e.g., BESS98\,\cite{BESS98}) separates the corrections into attenuation and production. Attenuation is due primarily to ionization and nuclear interactions and has a typical fractional deuteron reduction on the order of 8\%, regardless of energy. Atmospheric secondaries, dominated by production from proton and He primaries, have a much stronger energy dependence. The treatment of production in the literature is more varied—even the terminology used to treat a given process can vary significantly. In the BESS98 treatment, a method was derived\,\cite{BESS97} to use the secondary deuteron spectrum and the attenuation factor to calculate the TOA spectrum as a function of energy. The uncertainties propagated to the TOA spectra are on the order of 12\% for deuterons, and are due to the statistical uncertainty, the attenuation and secondary production uncertainties, the mass separation uncertainty, and the uncertainty due to effective geometric detector effects. Thus, it is important the uncertainty derived from atmospheric interactions be reduced in future balloon experiment measurements.

\section{The Propagation of CR Deuterons in the Galaxy and the Solar System}\label{sec3}

\subsection{Galactic propagation}
 
CRs propagate in the interstellar medium through a diffusive process attributed to interactions with the nonuniform Galactic magnetic field\,\cite{Blasi2013}. The microscopic interpretation of the CR diffusion is related to the ionized gas and the magnetic field produced by the particles in this plasma, which forms a magnetohydrodynamic (MHD) fluid. CRs are scattered in the interaction with these MHD waves, resulting in an effective diffusion movement\,\cite{crreview}. The mathematical formulation of the interaction between CRs and MHD waves results in the diffusion-reacceleration equation, an accepted model to describe CRs transport in the Galaxy. The general propagation equation for a particular CR species is\,\cite{crreview, Ptuskin2012}:

\begin{equation}\label{sec3:eq1}
\begin{aligned}
\frac{\partial \psi}{\partial t}=\mathbf{q}(\mathbf{r},p)+\nabla \cdot (D_{xx}\nabla \psi-\mathbf{V}\psi)+\frac{\partial}{\partial p}(p^2 D_{pp} \frac{\partial}{\partial p} \frac{\psi}{p^2}) \\ 
-\frac{\partial}{\partial p}[\psi\frac{\partial p}{\partial t}-\frac{p}{3}(\nabla \cdot \mathbf{V})\psi]-\frac{\psi}{\tau_f}-\frac{\psi}{\tau_\Gamma},
\end{aligned}
\end{equation}

\noindent where $\psi$ is the CR density per unit of momentum at position $\mathbf{r}$, and $\mathbf{q}(\mathbf{r},p)$ is the source term (which includes primary and secondary cosmic rays). The second term in Eq.\,(\ref{sec3:eq1}) represents the spatial diffusion process of CRs with $D_{xx}$ as the diffusion coefficient. $D_{xx}$ is, in general, a function of position ($\mathbf{r}$), velocity ($\beta$) and rigidity ($R$) of the particle. From microscopic theory, it is found that the diffusion coefficient is approximately $10^{28}\beta R_{\text{GV}}^{1/3} \text{\,cm}^2\text{\,s}^{-1}$ with a 1/3 exponent for a Kolgomorov-like turbulence\,\cite{crreview, gaggero_cosmic_2012}. However, it is not clear which turbulence spectrum dominates CR propagation or if a single spectrum is enough to explain CR diffusion in the whole energy range. Another type of turbulence spectrum, such as a Kraichnan-style spectrum with exponent 1/2, also provides good results and may be present in MHD turbulence. In general, the rigidity dependence of the diffusion coefficient is extracted from measurements, leaving the exponent as a parameter $\delta\approx0.3-0.6$. The third term accounts for the convective movement carried by Galactic winds transporting CRs with velocity $\mathbf{V}$. The next term in Eq.\,(\ref{sec3:eq1}) is the reacceleration term. This expression describes the stochastic acceleration experienced by CRs, with MHD waves being produced as a consequence of the movement of the medium. This, in turn, causes a change in the momentum of the particles. As can be seen in Eq.\,(\ref{sec3:eq1}), the reacceleration is interpreted as diffusion in momentum space with a diffusion coefficient $D_{pp}$ (subscript $pp$ means in momentum space). The coefficient is estimated as $D_{pp}=p^2V^2_{a}/9D_{xx}$, where $V_a$ is the Alfv\'en velocity\,\cite{crreview, Ptuskin2012}. The fourth term (${\partial p}/{\partial t}$) represents continuous energy losses in the Galactic disk from Coulomb collisions in an ionized medium and ionization losses. The fifth term ($\nabla \cdot \mathbf{V}$) describes adiabatic energy gains or losses resulting from nonuniform convection velocities whose inhomogeneities scatter CRs. $\tau_f$ is the timescale for losses by fragmentation or other processes like annihilation when antiparticles are propagated. $\tau_r$ is the timescale for radioactive decay\,\cite{crreview, Ptuskin2012}.

\subsection{Secondaries and the determination of propagation parameters}

The explicit role of secondary cosmic rays, including deuterons, in extracting information about CR propagation can be illustrated by using a simpler but meaningful approximation of Eq.\,(\ref{sec3:eq1}). In this approximation, energy loss, reacceleration, and convection contributions are neglected. The Galaxy is considered as a 2D system with CRs moving along the $z$ direction perpendicular to the Galactic disk. Following\,\cite{Jones_2001, Amato}, the solution for the secondary-to-primary ratio is:

\begin{equation}\label{sec3:eq2}
I_S/I_P = \frac{X(R)}{X_{P\rightarrow S}}\left[ 1+\frac{X(R)}{X_{S}} \right]^{-1},
\end{equation}

\noindent where $I_S$ and $I_P$ are the secondary and primary fluxes respectively, and $X(R) = \mu v H/2D(R)$ is the grammage of the ISM as a function of rigidity $R$, expressed in terms of the surface mass density of the Galactic disk $\mu$, the Galaxy halo height $H$, the particle velocity $v$, and the diffusion coefficient $D(R)$. $X_{P\rightarrow S}=m/\sigma_{P\rightarrow S}$ represents the critical grammage, i.e., the grammage necessary for a primary nucleus $P$ to break up into a secondary nucleus $S$ when interacting with the ISM, therefore, $\sigma_{P\rightarrow S}$ represents the fragmentation cross section from $P$ to $S$ and $m$ the mean mass of an interstellar particle. $X_{S}=m/\sigma_{S}$ is the critical grammage necessary for the secondary nucleus $S$ to interact with the ISM. Eq.\,(\ref{sec3:eq2}) is independent of any injection or primary component, allowing a direct view of the propagation element. For high rigidities, where $X(R) \ll X_{S}$, the secondary-to-primary ratio is proportional to $X(R)$, and the grammage as a function of rigidity can be estimated directly from measurements, i.e., the exponent $\delta$ in the diffusion coefficient ($D \propto \beta R^{\delta}$) can be extracted from a fit to data. This is the case for the B/C ratio that can be well fitted by a power-law function of rigidity above around 10\,GV. However, the interesting part to note for light nuclei is that the condition $X(R) \ll X_{S}$ is also met when the secondary nucleus has a long interaction mean free path or a low interaction cross section in the ISM, as in the case of deuterons. This implies that CR deuterons can provide access to propagation information without the limitation of being in the high-rigidity regime\,\cite{Vannuccini}. The fact that CR deuterons have a larger mean free path compared to heavier nuclei (on the order of twice larger), means they are less susceptible to suffering losses due to nuclear interactions with the ISM as in the case of heavier nuclei, and therefore they are more sensitive to the actual grammage $X(R)$ traversed by CRs in the Galaxy. How relevant and accessible a $d$/He ratio is in the light of new experiments and the impact this ratio has on updated propagation models will be explored in this work. 

\subsection{Solar modulation}

As cosmic rays approach the solar system, they encounter the heliosphere\,\cite{Zank}---a volume surrounding the Sun which is continuously filled with solar plasma. The boundary of this region is known as the heliopause.
The distribution of CRs is affected by interactions with the heliospheric magnetic field (HMF). The dominant component of the HMF is the so-called Parker spiral, generated by the rotation of the Sun. Irregularities in the HMF result in the diffusion of cosmic rays as they scatter off the irregularities\,\cite{Buchvarova}. A complete model of the heliosphere must therefore be capable of explaining the effects on CR distributions--namely, diffusion, drift, convection, and energy changes, described by the Parker transport equation\,\cite{Parker}:

\begin{equation}\label{sec3:eq6}
    \frac{\partial f}{\partial t}=-(\mathbf{V}+\langle \mathbf{V_{d} \rangle}) \cdot \nabla f + \nabla \cdot (\mathbf{K_{s}} \cdot \nabla f) + \frac{1}{3}(\nabla \cdot \mathbf{V}) \frac{\partial f}{\partial \mathrm{ln} R},
\end{equation}

where $f=f(\mathbf{r}, R, t)$ is a cosmic ray distribution function in a spherical heliocentric coordinate system with the equatorial plane oriented at a zenith angle of $90^\circ$, $\mathbf{V}$ is the solar wind velocity, $\langle \mathbf{V_d} \rangle$ is the time-averaged particle drift velocity, $\mathbf{K_s}$ is the diffusion tensor, and $R$ is the particle rigidity. The three terms on the right side of the equation represent convection, diffusion, and adiabatic energy changes, respectively. The solar modulation process over CRs is observed to have 11-year and 22-year cycles, the former being correlated to periods of extreme sunspot activity and the latter being due to the reversal of the heliospheric magnetic field at solar output extrema.

An approximate and widely-used solution to Eq.\,(\ref{sec3:eq6}) is the force-field approximation (FFA)\,\cite{GleesonAxford}. The FFA is a steady-state solution that relies on the assumptions of heliospheric spherical symmetry, the noninclusion of drift terms, a constant and radial solar velocity, and a boundary (heliopause) beyond which particles cannot travel, giving rise to the term ``force-field." The result of the approximation is a LIS multiplied (modulated) by an energy-dependent term, and the energies of cosmic rays are shifted down by a potential term known as the modulation potential. The FFA is a particularly good approximation for higher-energy cosmic rays. However, the noninclusion of drift terms, which become much more important at low energies and with increasing solar output, limits the application of this model. In addition, since it is a static solution, a single solution can typically only be applied over a few weeks due to the 11 and 22-year solar modulation cycles. These facts, combined with the modern availability of extensive computational resources, have resulted in the development of numerical solutions, like the one used in this work (see Sec.\,\ref{sec3:sub4}), that give a better description of the solar modulation process\,\cite{Potgieter}. 

\subsection{Simulation}\label{sec3:sub4}

Cosmic ray propagation in the Galaxy is treated in this work using the {\tt GALPROP\,v56}\,\cite{galprop, crreview} software package, which solves Eq.(\ref{sec3:eq1}) numerically. This analysis is based on the parameters obtained by Boschini \textit{et al.}\,\cite{Boschini_2020}, in which AMS-02, HEAO-3-C2, VOYAGER 1, and ACE-CRIS data were fit for nuclei ranging from H to Nickel (Ni) (not including H or He isotopes), in an energy range from 1\,MeV to 100\,TeV. In this model, the Galaxy is considered to be a cylinder with a halo height of $z=\text{4\,kpc}$ and a 20\,kpc radius. Moreover, the model uses a diffusion coefficient defined as $D_{xx} = D_{0}\beta^{\eta}(R/R_{0})^{\delta}$, where $\eta=0.7$ is a fitting parameter that improves agreement at low energies. This model also includes diffusive reacceleration, with Alfv\'en velocity $V_{a}$, and convection effects with a velocity gradient d$V$/d$z$. The values of the parameters are shown in Table\,\ref{sec3:tab1}, which were obtained through calibration with the most up-to-date data including the latest B/C results from AMS-02\,\cite{Boschini_2020}.

\begin{table}[t]
\begin{center}
\begin{tabular}{ c  c  c }                           
    Parameter & Best Value & Units \\
 \hline
 \hline
	$z$						& 4.0 	$\pm$ 0.6	  & $\text{kpc}$								\\
	$D_0/10^{28}$	& 4.3 	$\pm$ 0.7	  & $\text{cm}^2\text{s}^{-1}$	\\					
  $\delta$			& 0.415	$\pm$ 0.025 & -														\\
	$V_{a}$				& 30.0	$\pm$ 3.0   & $\text{km} \text{s}^{-1}$		\\   
  $dV/dz$				& 9.8 	$\pm$ 0.8   & $\text{km} \text{s}^{-1} \text{kpc}^{-1}$	\\
 \hline
\end{tabular}
\end{center}
\caption{Propagation parameter values and spectral parameters obtained from Reference\,\cite{Boschini_2020}.}
\label{sec3:tab1}
\end{table}

A full three-dimensional cosmic ray propagation model for describing CR modulation throughout the heliosphere is used to determine the differential intensity of cosmic ray deuterons at Earth. The modulation volume is assumed to be spherical with the heliopause position at 122 AU. This steady-state model is based on the numerical solution of Eq.\,(\ref{sec3:eq6}) and has been extensively described, and applied to interpreting observations of Galactic protons\,\cite{pot13}, electrons\,\cite{Vos_2015}, He\,\cite{Marcelli_2020} and positrons\,\cite{Aslam_2019}. The deuteron LIS obtained with {\tt GALPROP} as described above is speciﬁed as an initial condition in the numerical model. Other solar activity parameters are used in the modeling as inputs, such as the tilt angle $\alpha$, i.e. the angle that the dipole axis of the Sun forms with respect to the solar rotation axis, and the magnitude of the heliospheric magnetic field $B$ at the Earth, which changes continuously over time. In order to set up representative modulation conditions, appropriate moving averages of $\alpha$ and $B$ are used. The diffusion coefficients are tuned in order to reproduce a specific experimental dataset, and as such they are time-dependent. In this work, the diffusion coefficients used for the computation, as well as all the other relevant heliospheric parameters were the same as obtained for the PAMELA proton modeling described in\,\cite{pot13} and \,\cite{NGOBENI20222330}, that also reproduce AMS measurements of He isotopes.

\section{Deuteron formation in Cosmic Rays}\label{sec4}

As mentioned before, CR deuterons are produced mainly by the fragmentation of primary $^4$He and secondary $^3$He interacting with the ISM. An additional contribution of around 20\% comes from the fragmentation of heavier nuclei like C, N, and O\,\cite{Coste}. Since the ISM is mostly composed of H ($\sim$\,90\%) with a small contribution of He ($\sim$\,10\%), the releveant reactions in this CR analysis are $^4$He+$p$(He), $^3$He+$p$(He), and CNO+$p$(He). After a fragmentation interaction, the produced deuterons carry most of the projectile's energy, showing a peaked Gaussian distribution around it. Thus, as an acceptable approximation, the final deuteron kinetic energy per nucleon is considered to be similar in magnitude to the incoming He kinetic energy per nucleon\,\cite{Coste, Cucinotta}. Consequently, deuterons produced by fragmentation are dominant in the GeV region. Another important deuteron production process is the fusion of two protons through the reaction $p+p \rightarrow d+\pi^{+}$\,\cite{Coste, Meyer}. Although the $pp$ fusion reaction cross section is smaller than the one from He fragmentation, the proton flux is ten times more abundant than the He flux, making this contribution a large component of the final deuteron spectrum. Deuterons generated in the $pp$ fusion reaction contribute to the region below 1\,GeV, especially in the region between 80 and 250\,MeV/$n$\,\cite{Meyer}.

\begin{figure*}[t]
\begin{center}
\begin{tabular}{ll}
\includegraphics[width=7.6cm]{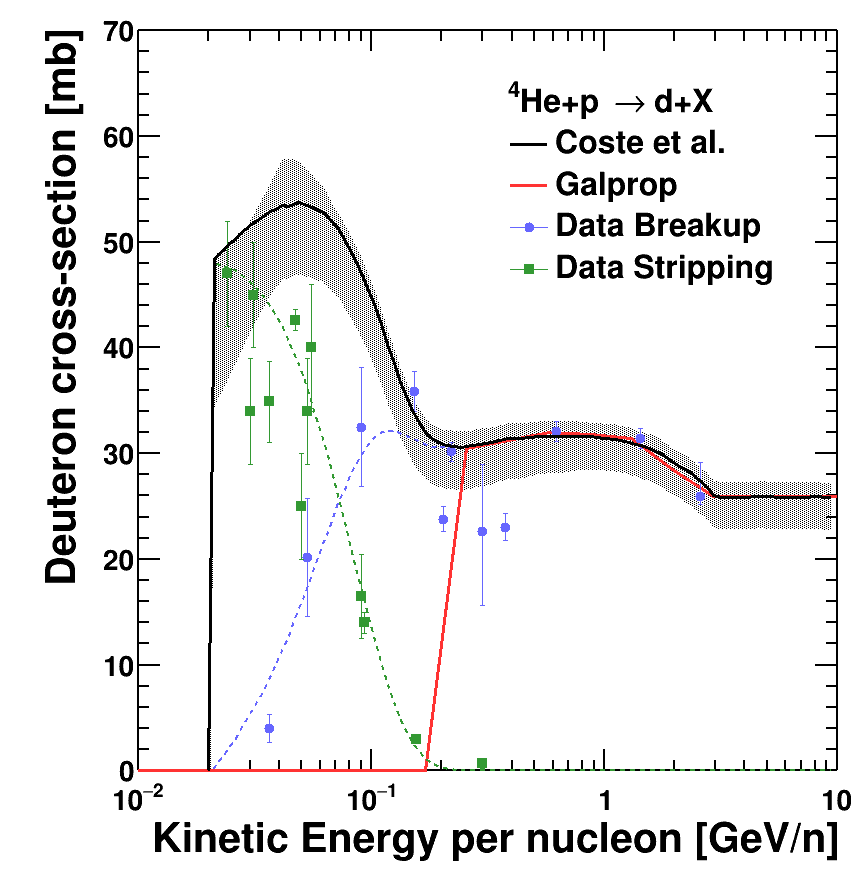}
& \includegraphics[width=7.6cm]{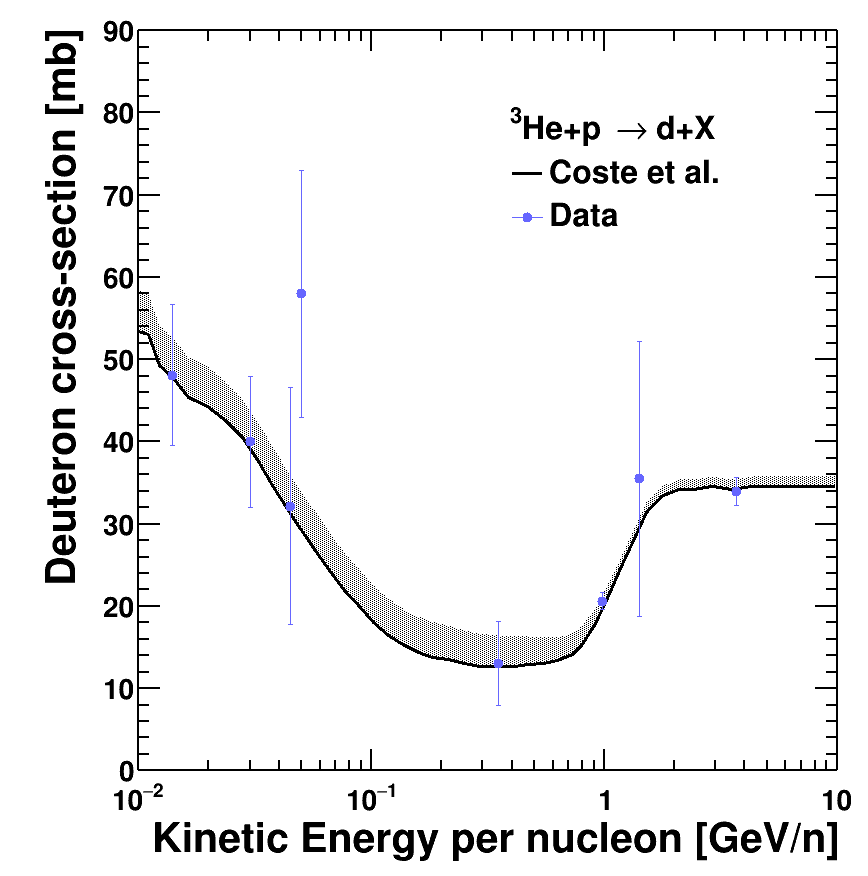}\\
\end{tabular}
\caption{Deuteron production cross sections: From $^{4}$He fragmentation (left plot) and from $^{3}$He fragmentation (right plot). Parametrizations from Coste \textit{et al.} and {\tt GALPROP} are compared to measurements (squares and circles)\,\cite{Meyer, Blinov&Chavedeya, Blinov, Jung, Griffiths, WebberCS, Nicholls, Glagolev}. More details are in the text.} 
\label{sec4:fig1}
\end{center}
\end{figure*}

Measurements of the reactions mentioned above are essential to model deuteron production cross sections at different energies and, therefore, to predict the deuteron flux correctly. An updated summary of these measurements from accelerator experiments as of 2008 is presented in Coste \textit{et al.}\,\cite{Coste}  (see Appendix B in\,\cite{Coste}) and revised in this paper (see Fig.\,\ref{sec4:fig1}). In the work by Coste \textit{et al.}, an improved parametrization of the cross sections based on the works by Cucinotta \textit{et al.}\,\cite{Cucinotta} and Meyer\,\cite{Meyer} is shown. This parametrization is the result of two contributions: a pickup process dominating below 0.1\,GeV/$n$, where the incident proton tears a neutron or proton off the He nuclei, and a break-up process ruling above 0.1\,GeV/$n$, where the energy is enough to dissociate the incoming He in nucleons that can form new light nuclei by coalescence. In Fig.\,\ref{sec4:fig1}, the data samples\,\cite{Meyer, Blinov&Chavedeya, Blinov, Jung, Griffiths, WebberCS, Nicholls, Glagolev} and parametrizations used by Coste \textit{et al.} are presented for $^4$He+$p$ and $^3$He+$p$ fragmentation reactions as a function of kinetic energy per nucleon. In the case of the $^4$He+$p$ reaction (Fig.\,\ref{sec4:fig1} left panel), the parametrization for pickup and break-up contributions are shown as green and blue dashed lines respectively. Likewise, green squares and blue circles data points represent measurements for these two processes accordingly. The sum of these two contributions results in the total production cross section shown as a solid black line. For the $^3$He+$p$ reaction (Fig.\,\ref{sec4:fig1} right panel), only the total deuteron production cross section is shown along with available measurements. 

To obtain a correct prediction of deuteron CRs in the Galaxy, all cross sections from reactions related to deuteron production must be included in Eq.\,(\ref{sec3:eq1}). The model used in {\tt GALPROP} for deuteron production cross section in $^4$He+$p$ interactions can be seen in the left panel of Fig.\,\ref{sec4:fig1} (solid red line). {\tt GALPROP's} parameterization of the cross section is based only on the most recent data by\,\cite{Blinov&Chavedeya} for energies above 0.2\,GeV/$n$. Furthermore, {\tt GALPROP} does not include a deuteron production cross section from the $^3$He+$p$ reaction. Therefore, the first goal of this work was to implement updated parametrizations of the deuteron production cross sections developed by Coste \textit{et al.} in {\tt GALPROP} (a similar study was made by\,\cite{Picot2015}). From differences between the red and black solid lines in Fig.\,\ref{sec4:fig1}, it is foreseeable that there will be differences in the final deuteron flux when these cross sections are replaced. For example, an increase at low energies due to the inclusion of the stripping component in the cross section and an increase at energies above 1\,GeV/$n$ due to the inclusion of the $^{3}$He fragmentation. Parametrizations for $^4$He+He and $^{3}$He+He reactions were also included in {\tt GALPROP} based on Coste et al results. The $pp$ fusion reaction was added to {\tt GALPROP\,v56} taking the implementation made by \cite{Picot2015} to a previous version of the official code. This implementation was based on the work by Meyer\,\cite{Meyer}. Heavier nuclei fragmentation cross sections resulting in deuterons were also updated for this study. The contribution from the groups CNO, MgAlSi, and FeNi were implemented in {\tt GALPROP} using results from Coste \textit{et al.} (see Appendix\,\ref{app:a:sb1}).

\begin{figure*}[t]
\centering
\includegraphics[width=0.47\linewidth]{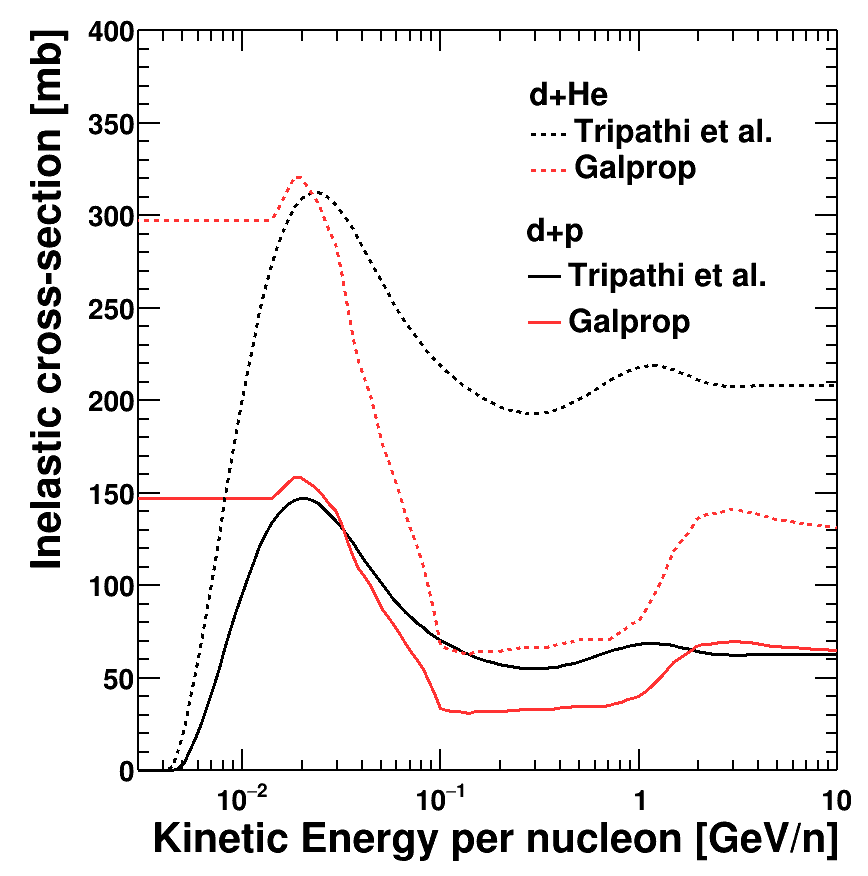}
\caption{Deuteron total inelastic cross sections: The parameterization by Tripathi \textit{et al.}\,\cite{Tripathi}, used in this work, is compared to the parameterization generally used with {\tt GALPROP}\,\cite{Barashenkov}, for $d$+H and $d$+He interactions.}
\label{sec4:fig2}
\end{figure*}

An essential component in calculating the deuteron production in CRs is the uncertainty related to the production cross section. This uncertainty can be seen in Fig.\,\ref{sec4:fig1} as a gray band in both reactions, and it was estimated based on residuals between parametrization and available data, weighted by the errors reported and considering an approximated systematic error of 5\% for the most accurate dataset\,\cite{Blinov&Chavedeya}. Due to an asymmetric distribution of measurements above and below the parameterization, the calculated uncertainty is also asymmetric. In this work, as in the previous work by\,\cite{Tomassetti_2012} this uncertainty is introduced in the propagation process for deuterons. Note that the measurement at the highest energy ($\sim$3\,GeV/$n$) in the $^4$He+$p$ reaction (Fig.\,\ref{sec4:fig1} left panel) is below measurements by the same experiment at lower energies (0.5-2\,GeV/$n$). This can be a consequence of undercounting some of the reactions producing deuterons as final states\,\cite{Blinov&Chavedeya, Glagolev, Coste}, therefore, a higher error band has been estimated. Since this measurement is important to define the cross section for energies above 3\,GeV/$n$, an alternative scenario without this point is analyzed in Sec.\,\ref{sec5}.

After production, deuterons undergo a series of inelastic interactions with the ISM, reducing the number that arrives at Earth. Total inelastic cross sections of light nuclei with H and He targets are described by Tripathi \textit{et al.}\,\cite{Tripathi} using a parametric model from the MeV to GeV energy range. Tripathi's parametrization has been successful in describing measurements and is used widely in different areas related to the transportation of ions in matter\,\cite{AGOSTINELLI2003250}. Differences between the model considered in {\tt GALPROP} and Tripathi's prediction for $d$+He and $d$+$p$ interactions are shown in Fig\,\ref{sec4:fig2}. The reason for these differences is that {\tt GALPROP} uses a general parameterization for $p$+A reactions by Barashenkov and Polanski\,\cite{Barashenkov}, while Tripathi's parameterization is specific for light-nuclei interactions. Hence, in this work, Tripathi's parametrization was implemented in {\tt GALPROP} for a better description of the deuteron total inelastic cross sections. All these improvements to the deuteron flux prediction become crucial when comparing to present and future high-precision measurements.


\section{Results for Cosmic Ray Deuterons}\label{sec5}

\begin{figure}[t]
\centering
\begin{tabular}{ll}
\includegraphics[width=0.47\linewidth]{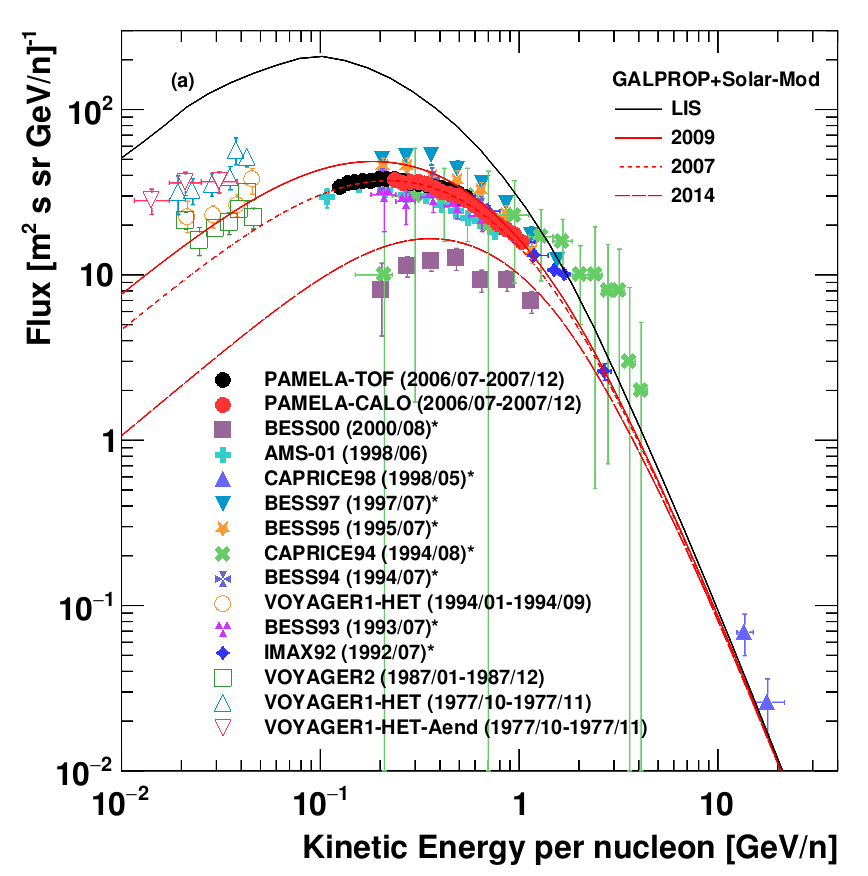}
& \includegraphics[width=0.47\linewidth]{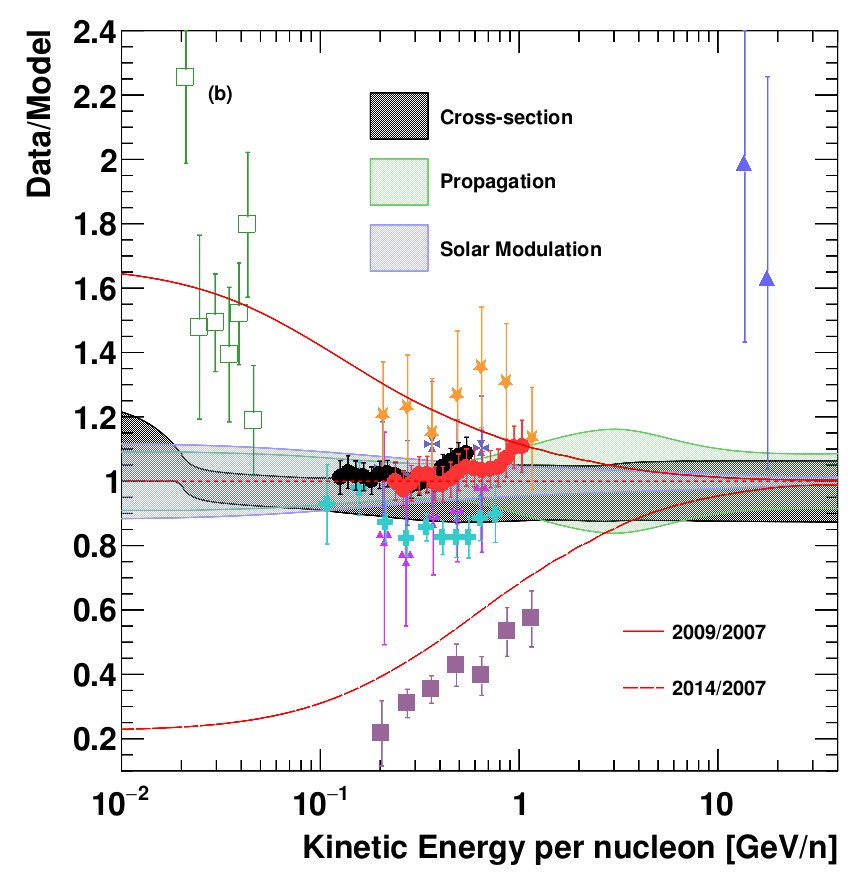} \\
\multicolumn{2}{c}{\includegraphics[width=0.47\linewidth]{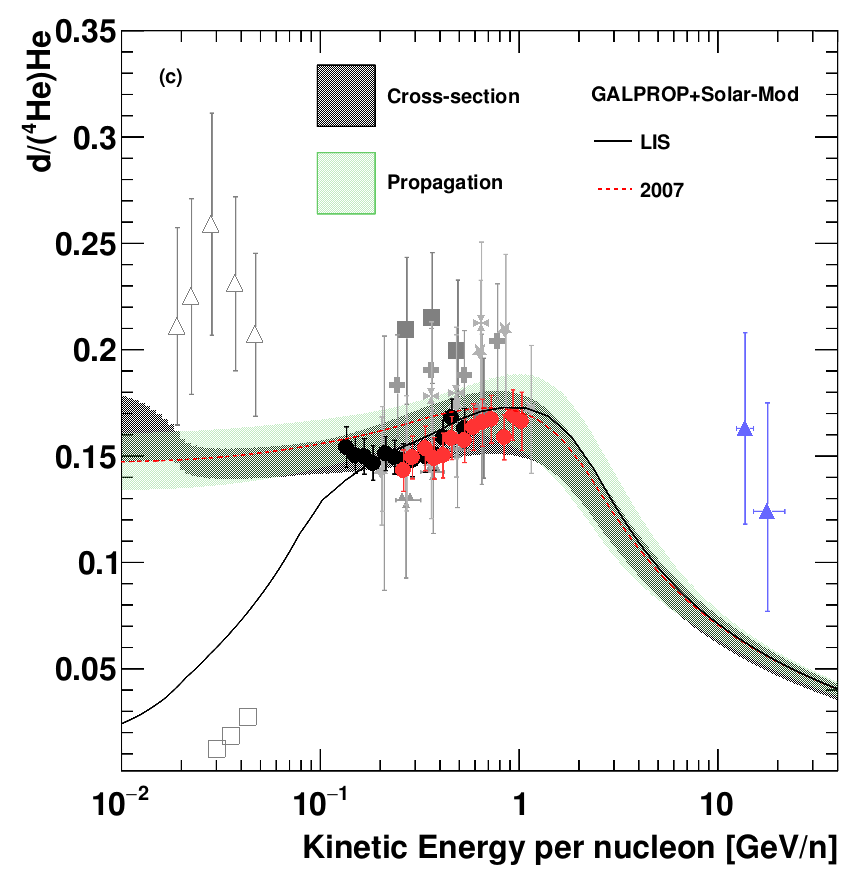}}
\end{tabular}
\caption{ Deuteron results: Plot\,(a) shows the deuteron LIS obtained with {\tt GALPROP} along with the deuteron spectra for three solar modulation periods obtained with the 3D numerical model corresponding to 2007, 2009, and 2014. Plot\,(b) shows the flux ratio between data and simulation for the 2007 solar period. Uncertainties derived from production cross section, propagation, and solar modulation are included. Plot\,(c) shows the deuteron over $^4$He ratio simulation for the 2007 solar period compared to $d$/$^{4}$He and $d$/He($^{*}$) measurements.}
\label{sec5:fig1}
\end{figure}

After introducing the updated deuteron production cross sections described in Sec.\,\ref{sec4} into the {\tt GALPROP} propagation model, and using the propagation parameters listed in Table\,\ref{sec3:tab1}, the LIS deuteron flux was calculated (see solid black line in Fig.\,\ref{sec5:fig1}\,(a)). The effect of the solar activity over this new deuteron LIS flux was modeled as described in Sec.\,\ref{sec3}. The modulated deuteron spectrum averaged over a period of one year was obtained for three different periods of time. The three modulated spectra are representative of different periods of solar activity: 2007 (dotted red line), which was a period in between a maximum and a minimum of solar activity, 2009 (continuous red line), which was the minimum of the solar cycle 23, and 2014 (dashed red line), which corresponds to the period of the maximum of solar cycle 24.\,\cite{Fu_2021}. The results from these simulations are compared to the available data as shown in Fig.\,\ref{sec5:fig1}\,(a). In Fig.\,\ref{sec5:fig1}\,(b), the data are compared to the model by calculating the ratio for every data sample to the predicted flux in 2007 (dashed red line). The model results in 2009 (solid red line) and 2014 (long dashed red line) were also divided by the flux from 2007. In this figure, shadow bands with different colors have been plotted to represent uncertainties from cross section (gray), propagation (green), and solar modulation (blue). The cross section uncertainty was derived from the study in Sec.\,\ref{sec4}, estimating the deviation between the best-fit model and accelerator data (see Fig.\,\ref{sec4:fig1}). This asymmetric uncertainty was propagated to the flux calculation with {\tt GALPROP}, obtaining a deviation of around 10\% below the mean value of the parameterization for most of the energy range. In the low-energy region ($<$0.1\,GeV/$n$) where the cross section is dominated by the stripping process the uncertainty increases in the flux as a consequence of the higher errors in accelerator measurements. The propagation uncertainty was obtained by varying the best-fit values of the propagation parameters between the minimum and maximum errors reported in Table\,\ref{sec3:tab1}, and calculating the deviation of the fluxes associated with those parameters. As observed in Fig.\,\ref{sec5:fig1}\,(b), the propagation uncertainty is around 10\% below 1\,GeV/$n$, but increases to 20\% around 3\,GeV/$n$, and decreases to about 10\% above 10\,GeV/$n$. The bump at 3\,GeV/$n$ is mostly due to the variation of the normalization parameter in the spatial diffusion coefficient $D_{0}$ and the Alfv\'{e}n velocity as part of the reacceleration process. When the diffusion coefficient decreases or the Alfv\'{e}n velocity increases, the deuteron flux is redistributed in energy, increasing its value in the energy region above 3\,GeV/$n$ and reducing its value below this energy. An opposite result is obtained when the diffusion coefficient increases or the Alfv\'en velocity decreases. The uncertainty associated with the solar modulation model was derived by calculating the deviation between the model and PAMELA measurements for He in similar periods. An approximate 10\% difference is observed below 1\,GeV/$n$, with a decreasing effect at higher energies, as expected.

The first important observation in Fig.\,\ref{sec5:fig1}\,(b) is the good agreement between the simulation result for 2007 (dashed red line) and PAMELA data. As can be seen, measurements are well within model uncertainties. The theoretical prediction for this period of medium solar activity is also close to AMS-01, BESS93, and BESS94, consistent with low-medium solar activity periods between solar cycles 22 and 23. The result for the 2014 period (long dashed red line) is close to BESS00 data and is compatible with a high solar activity time in cycle 23. In the case of 2014 prediction, the overproduction by the model could be explained by a lower intensity in solar cycle 24 (2009) than in cycle 23 (2000). However, a dedicated analysis is outside the scope of this work. Additionally, the flux in 2009 (solid red line) corresponding to the solar minimum between cycles 23 and 24 is close to the BESS95 and BESS97 measurements taken during a solar minimum activity between cycles 22 and 23. This result in 2009 is also close to the data from VOYAGER\,1 and 2 taken during solar minimum in cycles 20, 21, and 22. It is essential to clarify that since solar cycles have different intensities, it is not expected that the simulation calculated in this work for cycle 24 entirely matches measurements taken in different solar cycles. 

In Fig.\,\ref{sec5:fig1}\,(c), the simulation result for deuterons over $^{4}$He is presented in comparison to $d$/$^{4}\text{He}$ and $d$/He measurements. As can be seen, the model describes PAMELA, BESS93, and BESS94 data within the model uncertainties (although it tends to overestimate PAMELA measurements). The result is also close to AMS-01 data and other BESS results that have uncertainties on the order of 20\%. This result is remarkably satisfactory, considering none of the data shown in the plot were used to tune the model, and that the prediction is based entirely on B/C, $p$, and He flux data. At energies below 0.1\,GeV/$n$, the model underpredicts VOYAGER\,1 measurement by around 30\%---as expected from what was observed in the deuteron flux. For the case of VOYAGER\,2, measurements are closer to the LIS prediction, which is explained by an anomalously large $^{4}$He measurement\,\cite{Voyager2}. At higher energies, the result from this work is well below CAPRICE98 measurements, for example, the difference between the lower error limit for the data point at 17.8 GeV/$n$ and the model at that energy is around 30\% (see Fig.\,\ref{sec5:fig1}\,(c)), with an uncertainty in the model from cross section on the order of 6\%. If the data point at 3\,GeV/$n$ in the left panel of Fig.\,\ref{sec4:fig1} is not included due to its large error, the predicted value of the model for the $d$/$^{4}\text{He}$ ratio could increase to about 10 to 15\%, still below the lower limit measured by CAPRICE98. This could indicate a discrepancy in the diffusion model for light nuclei compared to B/C. However, given the magnitude of the uncertainties of these measurements (order of 30\%-40\%), it is difficult to reach a conclusion.

\begin{figure}[t]
\centering
\includegraphics[width=0.5\linewidth]{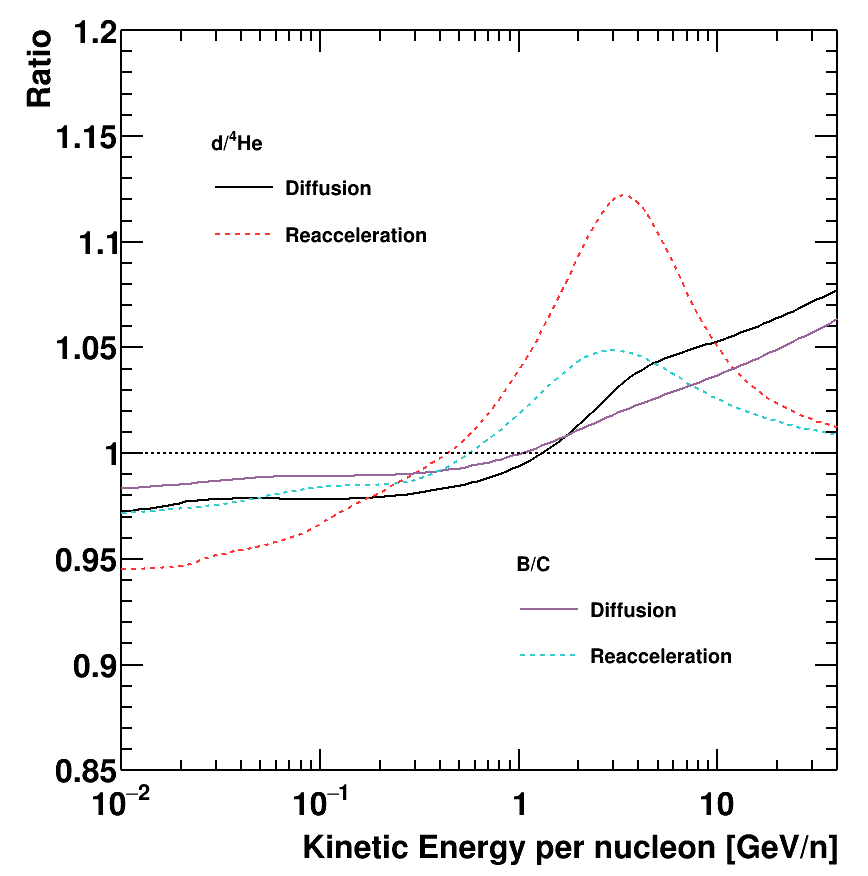}
\caption{${\boldsymbol d}$/$^4$He compared to B/C: The ratio between results before and after perturbation of the reacceleration and diffusion parameters for $d$/$^4$He and B/C.}
\label{sec5:fig3}
\end{figure}

\begin{figure*}[t]
\centering
\begin{tabular}{ll}
\includegraphics[width=0.47\linewidth]{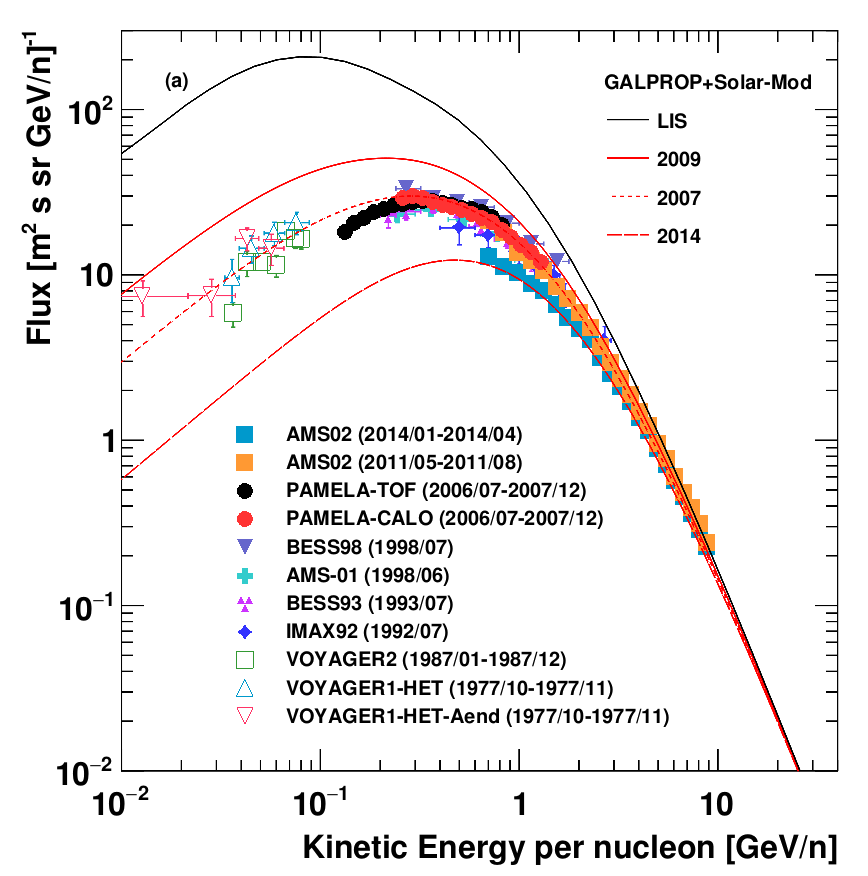}
& \includegraphics[width=0.47\linewidth]{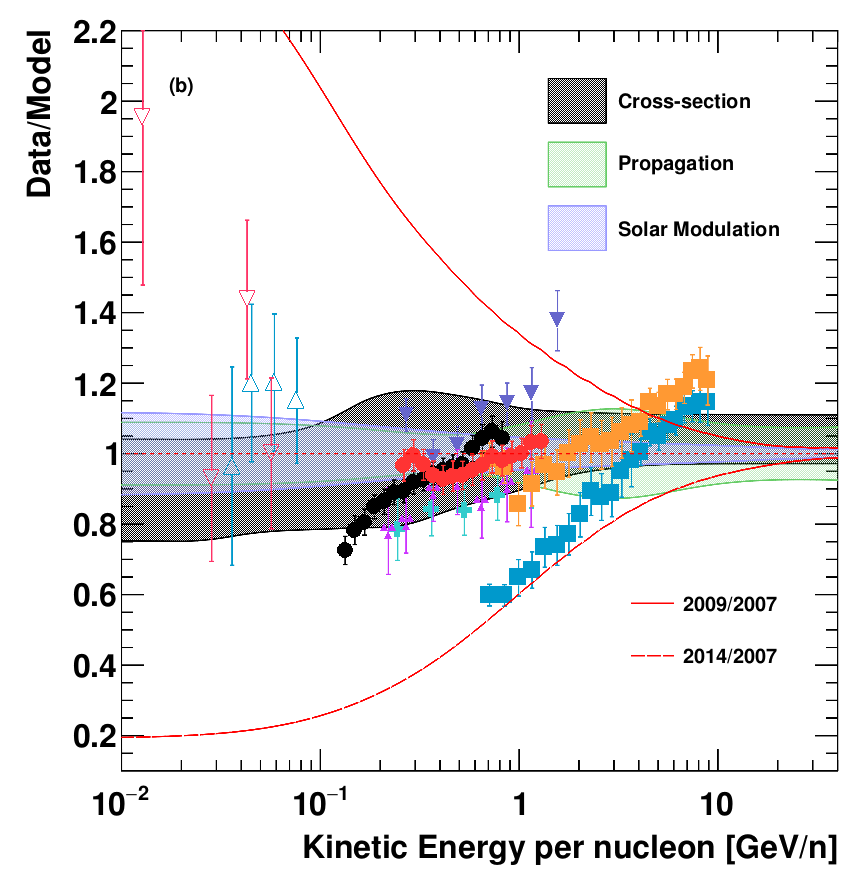}\\
\multicolumn{2}{c}{\includegraphics[width=0.47\linewidth]{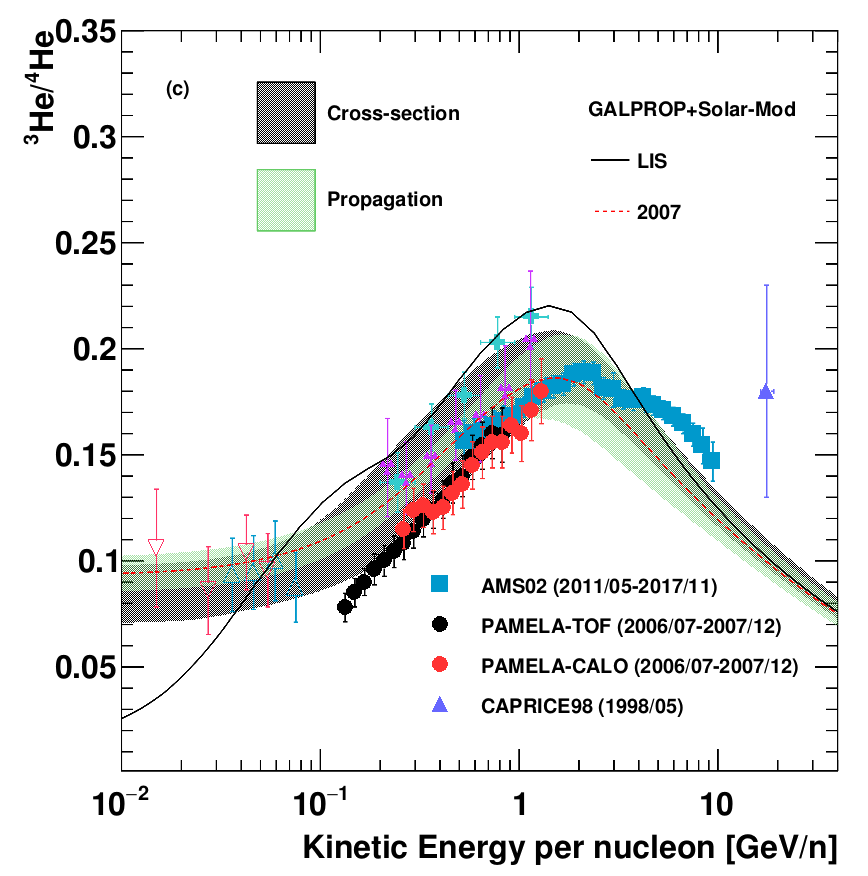}}
\end{tabular}
\caption{$^{3}$He results: Plot\,(a) shows flux measurements and simulation results for LIS and three different periods using {\tt GALPROP} and the 3D solar modulation model. Plot\,(b) shows the data-over-model ratio, including uncertainties in the model from cross sections, propagation parameters, and solar modulation. Plot\,(c) shows the $^{3}$He/$^{4}$He ratio simulation result with uncertainties compared to data.}
\label{sec5:fig2}
\end{figure*}

In general, it is observed that in $d$/$^4$He there is a crucial energy region from 1 to 20~GeV/$n$ which lacks precision measurements. Here, the influence of solar modulation is diminished, and the slope of $d$/$^4$He is influenced by two main propagation processes: diffusion and reacceleration. To illustrate their effects, the exponent index $\delta$ in the diffusion coefficient ($D_{xx}=D_{0}\beta^{\eta}(R/R_{0})^{\delta}$), and the Alv\'en velocity $V_{a}$ in the reacceleration term are perturbed separately in the model when the flux is calculated. Then, the resulting flux is compared to the initial one, as shown in Fig.\,\ref{sec5:fig3}. A similar calculation has been made for B/C. As can be seen, reacceleration strongly affects $d$/$^4$He and B/C ratios, redistributing the flux from lower energy regions, where cosmic rays spend more time traveling in the Galaxy and interacting with MHD waves, to a higher energy region above 1\,GeV/$n$. The reduction of the reacceleration intensity at high energies ($\sim$\,10\,GeV/$n$) and the asymmetry observed at the tail is a consequence of the coupling between diffusion and reacceleration ($D_{pp}=p^2V^2_{a}/9D_{xx}$). At high energies, diffusion increases, making the reacceleration energy-gaining process less efficient. Moreover, reacceleration and diffusion have a more significant effect on $d$/$^4$He than B/C for high energies, which is a direct consequence of light nuclei having a lower interaction cross sections with the ISM than heavier nuclei. As seen in Fig.\,\ref{sec5:fig3}, the diffusion in $d$/$^4$He increases with energy, it starts to rise above 1\,GeV/$n$ and dominates over reacceleration above 10\,GeV/$n$. Such characteristics highlight the significance of $d$/$^4$He measurements above 1\,GeV/$n$ to constrain model parameters, such as the exponent in the diffusion coefficient. 
underestimating some accelerator measurements that possibly did not account for all reactions producing these light nuclei
Additionally, as a test of consistency, the $^3$He flux and the $^3$He/$^4$He ratio were calculated following a similar approximation as for deuterons. The results are compared to measurements and presented in Fig.\,\ref{sec5:fig2}. The $^3$He flux for 2007 agrees with PAMELA data, the description is better for the calorimeter part of the data than for TOF, but measurements are within model uncertainties in most of the energy range. The dominant uncertainty in the prediction is from cross sections, particularly in the region from 0.1 to 1.0\,GeV/$n$, where discrepancies from different data samples in accelerator experiments and important uncertainties in the measurements affect the parameterization (see Appendix\,\ref{app:a:sb2}). The agreement between simulation and AMS-02 data for the 2014 period is good for energies below $\sim$\,3-4\,GeV/$n$, but the model starts underpredicting the $^3$He flux above these energies. However, the deviation between model and data does not exceed the model uncertainty. The $^3$He/$^4$He ratio in Fig.\,\ref{sec5:fig2}\,(c) shows that the model describes PAMELA and AMS-02 data within cross section uncertainties for most of their energy ranges. Nevertheless, the model underpredicts AMS-02 data above $\sim$\,5-6\,GeV/$n$ by less than 10\%, and again as for the case of deuterons, the result is well below CAPRICE98 measurements. Results for $^4$He flux can be seen in Appendix\,\ref{app:b}.

\section{Conclusions}\label{sec6}

In this work, CR deuterons have been reviewed by studying the mechanisms for their production in CRs interactions, their propagation in the Galaxy, and the effect of solar modulation on them, highlighting the important role they play in deciphering cosmic ray diffusion properties in the Galaxy.

A state-of-the-art calculation of the deuteron flux and $d$/$^4$He was performed using {\tt GALPROP}\,v56 as a propagation modeling tool, and considering a diffusion-reacceleration model with the Boschini \textit{et al.}\,\cite{Boschini_2020} parameters obtained by fitting AMS-02, HEAO-3-C2, VOYAGER 1, and ACE-CRIS data for nuclei from H to nickel and the B/C ratio. Deuteron and $^{3}$He production cross sections were updated in {\tt GALPROP} by using the latest parametrizations from accelerator experiments \,\cite{Coste}, and the related uncertainties were included in the calculations. It was found that this uncertainty is less than 10\% for deuterons and below 20\% for $^{3}$He in most of the energy range. These uncertainties result from underestimating some accelerator measurements that possibly did not account for all reactions producing these light nuclei. Especial attention has been paid to cross-section uncertainties at $\sim$ 3\,GeV/$n$ that have an impact on the model prediction for the energy region where AMS-02 measurements are expected. Additionally, cross-section uncertainties associated to Boron and other secondary species, also introduce uncertainties in the propagation parameters\,\cite{De_La_Torre_Luque_2022}. This stresses the need for better measurements of nuclear reactions in energy regions of interest for galactic CRs. 

It was also found that the simulation results describe the deuteron flux and $d$/$^4$He data below 1\,GeV/$n$ within the uncertainties of the model, and show consistency with $^3$He/$^4$He ratio. This result shows the good prediction power of {\tt GALPROP} and Boschini \textit{et al.}, propagation parameters after introducing all necessary deuteron and $^3$He cross sections. However, the simulation underestimates the only measurements available at high energy by CAPRICE98 for $d$/$^4$He and $^3$He/$^4$He ratios, characterized by their large uncertainties. To accommodate such a difference between the model and data, it would be necessary to modify the parameters related to diffusion and reacceleration for deuterons and $^3$He (secondary light species) with respect to those obtained with heavier nuclei (B/C). If such a difference is corroborated by future high-precision measurements in CRs, and reduced uncertainties in the model, a general calibration of the propagation parameters using B/C ratio would no longer be valid, and could imply a break in the universality of cosmic ray propagation. Further studies and high-precision measurements of $d$/$^4$He for energies above 1\,GeV/$n$ are essential to arrive at a definite conclusion.

The above results further demonstrate that the observed reacceleration, generally considered necessary to explain B/C around 1\,GeV/$n$, has a higher impact on $d$/$^4$He and $^3$He/$^4$He over the energy region between 1 and 10\,GeV/$n$, and this effect decreases rapidly above that energy. This is expected because of the lower interaction cross section for light nuclei. This result is in agreement with other works\,\cite{WU2019292,Weinrich} where diffusion-reacceleration models produced the best fits to data when $^3$He or $d$/$^4$He measurements were included, despite estimating different parameter values. Therefore, both the effects of reacceleration and diffusion would be strongly constrained by new high-precision deuteron and $d$/$^4$He measurements in the energy range from 1 to 10\,GeV/$n$. 

A new version of {\tt GALPROP (v57)} was recently released\,\cite{galpropv57} during the preparation of this work, and it includes some of the improvements in cross sections presented in Sec.\ref{sec4}. The implementations of these cross sections in this work are in agreement with those in {\tt GALPROP v57}.


\section{Acknowledgments}\label{sec7}

D.\,Gomez-Coral and P.\,von Doetinchem would like to thank the National Science Foundation for supporting this work under the award No.\,PHY-2013228. The contributions of collaborator C. Gerrity are supported by the NASA FINESST award No.\,80NSSC19K1425. 

\newpage
\appendix

\section{CROSS SECTIONS}\label{app:a}

\subsection{CNO, MgAlSi, and FeNi fragmentation cross sections}\label{app:a:sb1}

\begin{figure*}[t]
\centering
\begin{tabular}{ll}
\includegraphics[width=0.47\linewidth]{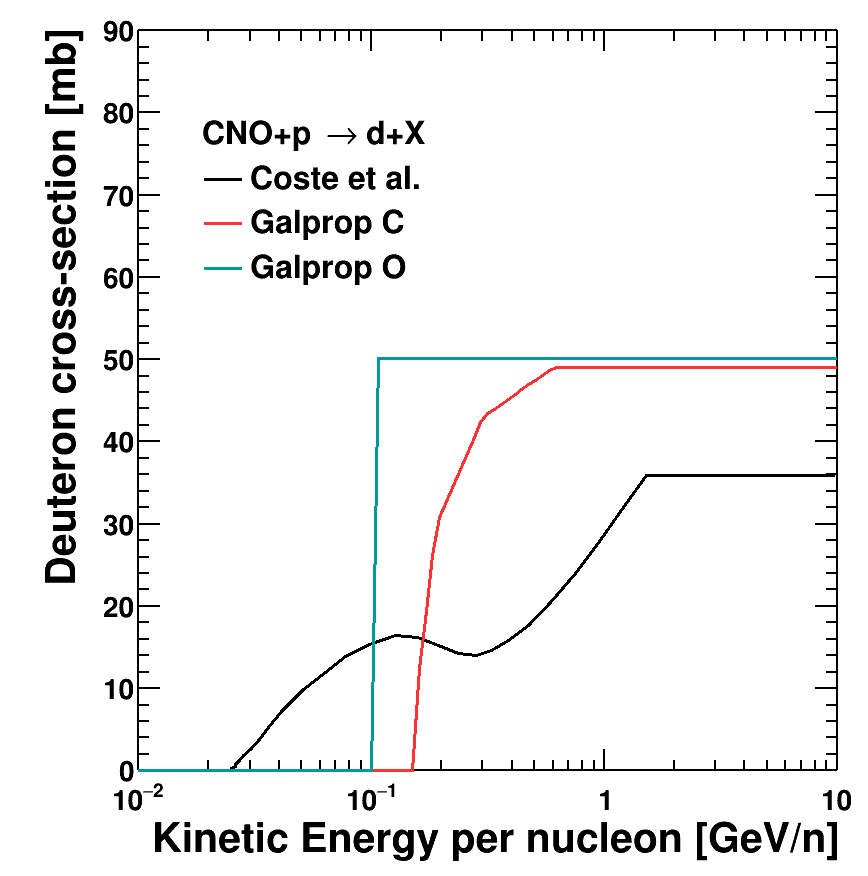}
& \includegraphics[width=0.47\linewidth]{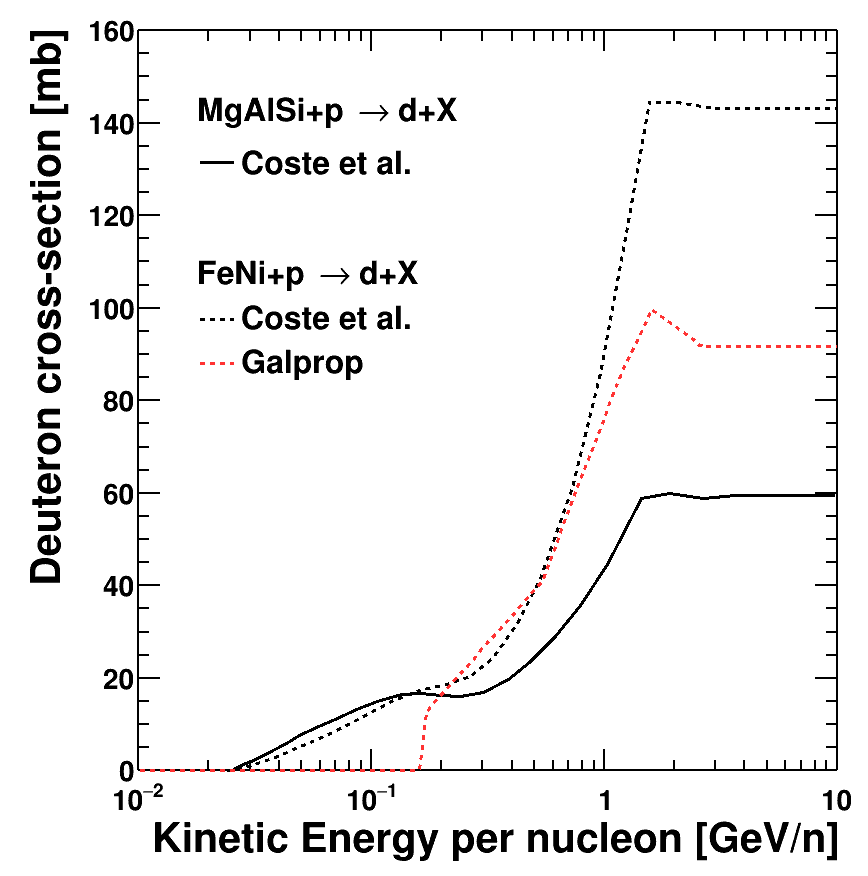}
\end{tabular}
\caption{Deuteron production cross sections: Parametrizations by Coste \textit{et al.} and {\tt GALPROP} for projectiles with A$>$4 interacting on proton target.}
\label{a1:fig1}
\end{figure*}

In Fig.\,\ref{a1:fig1}, parametrizations from Coste \textit{et al.}\,\cite{Coste} describing the production cross section of deuterons from heavier nuclei colliding with a proton target are presented. Based on an extended set of measurements in the energy range below 10\,GeV/$n$, these parametrizations are separated into three groups of projectile nuclei: CNO, MgAlSi, and FeNi which have similar cross sections. The models for deuteron production from C, O, and Fe fragmentation followed originally by {\tt GALPROP} are included for comparison in the same plot. {\tt GALPROP\,v56} does not include deuteron production cross sections from N, Mg, Al, Si, or Ni.

\subsection{$^{3}$He cross sections}\label{app:a:sb2}

\begin{figure*}[t]
\centering
\begin{tabular}{ll}
\includegraphics[width=0.47\linewidth]{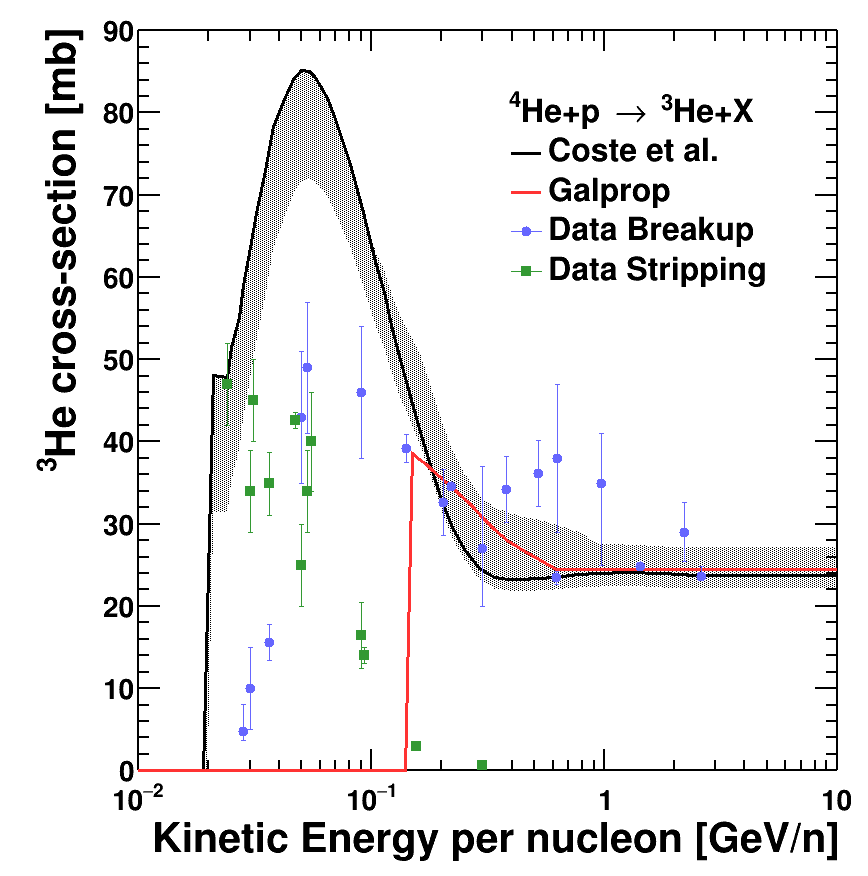}
& \includegraphics[width=0.47\linewidth]{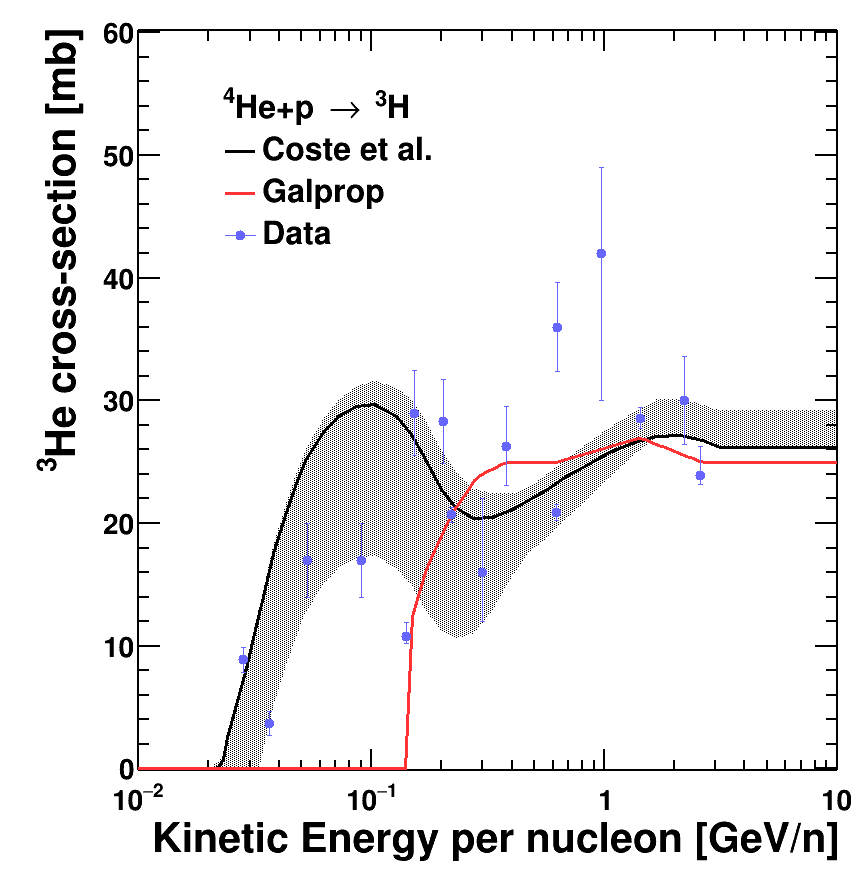}\\
\multicolumn{2}{c}{\includegraphics[width=0.47\linewidth]{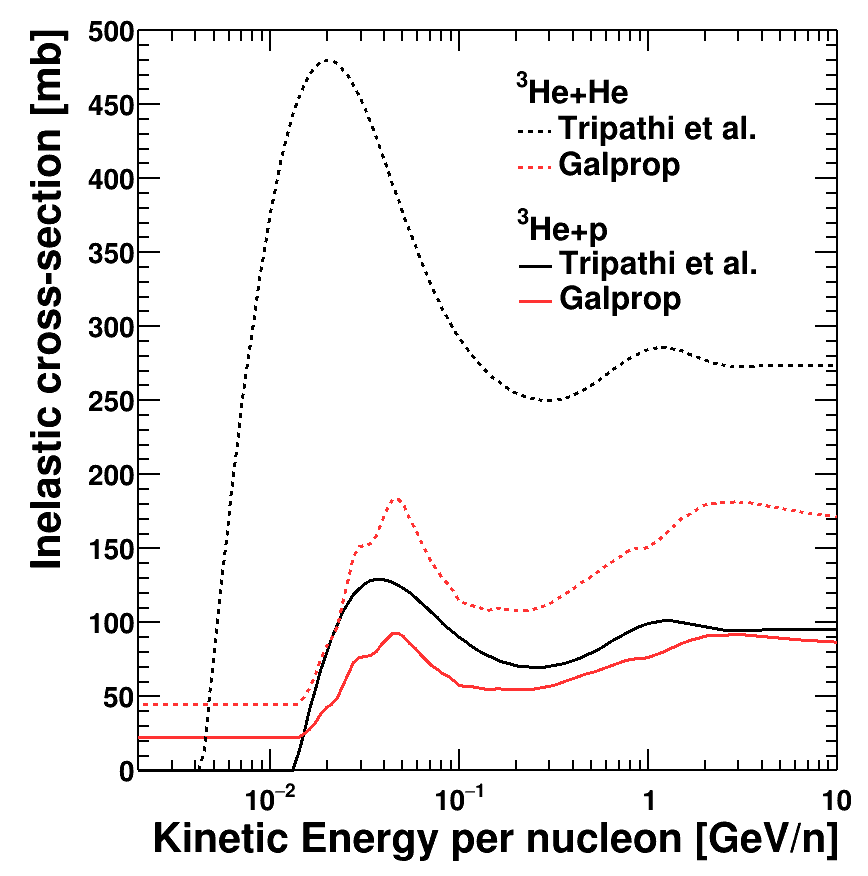}}
\end{tabular}
\caption{$^{3}$He cross sections: Production cross section from $^4$He fragmentation (left). Right: Tritium production cross section from $^4$He fragmentation}
\label{a1:fig2}
\end{figure*}

CR $^3$He is also created mainly from $^4$He fragmentation with the ISM, as in the case of deuterons. However, $^3$He is also produced through the decay of tritium created from $^4$He spallation. The top-left panel in Fig.\,\ref{a1:fig2} shows a compilation of $^3$He production measurements collected by Coste \textit{et al.}\,\cite{Coste}, based on data samples by\,\cite{Meyer, Blinov&Chavedeya, Blinov, Jung, Griffiths, WebberCS, Nicholls, Glagolev}. As in deuteron formation, two main processes are involved in $^3$He production from $^4$He, stripping at low energy and break-up at higher energies. Additionally, the improved parameterization developed by Coste \textit{et al.} is included with an estimation of the uncertainty based on errors from the data points presented. A comparison to the {\tt GALPROP} model shows it agrees with the new parameterization above 0.1\,GeV/c, since both are based on the same measurements. Still, it also shows the limitation of {\tt GALPROP} to describe the stripping contribution below 0.1\,GeV/c. The top-right panel in Fig.\,\ref{a1:fig2} shows the production of triton from $^4$He interaction with protons. In this case, only the break-up process is present, and measurements from different experiments, including those based on n-$^4$He reactions, are compared to the model. The uncertainty has been estimated based on the prediction by the parameterization and measurement errors. The figure also shows the {\tt GALPROP} model, which differs at lower energies because it is based on recent data by\,\cite{Blinov&Chavedeya}, without including all measurements in the plot. Another critical component that was changed in {\tt GALPROP} was the total inelastic cross sections of $^{3}$He when interacting with protons and He. The bottom panel in Fig.\,\ref{a1:fig2} shows the parametrizations by Tripathi \textit{et al.}\,\cite{Tripathi}, which are based on measurements and have been explicitly calculated for light nuclei, in comparison to the original {\tt GALPROP} parametrizations. It is important to note the significant difference in cross sections for $^{3}$He concerning a He target affects $^{3}$He fluxes.

\section{$^{4}$He RESULTS}\label{app:b}

\begin{figure*}[t]
\centering
\begin{tabular}{ll}
\includegraphics[width=0.47\linewidth]{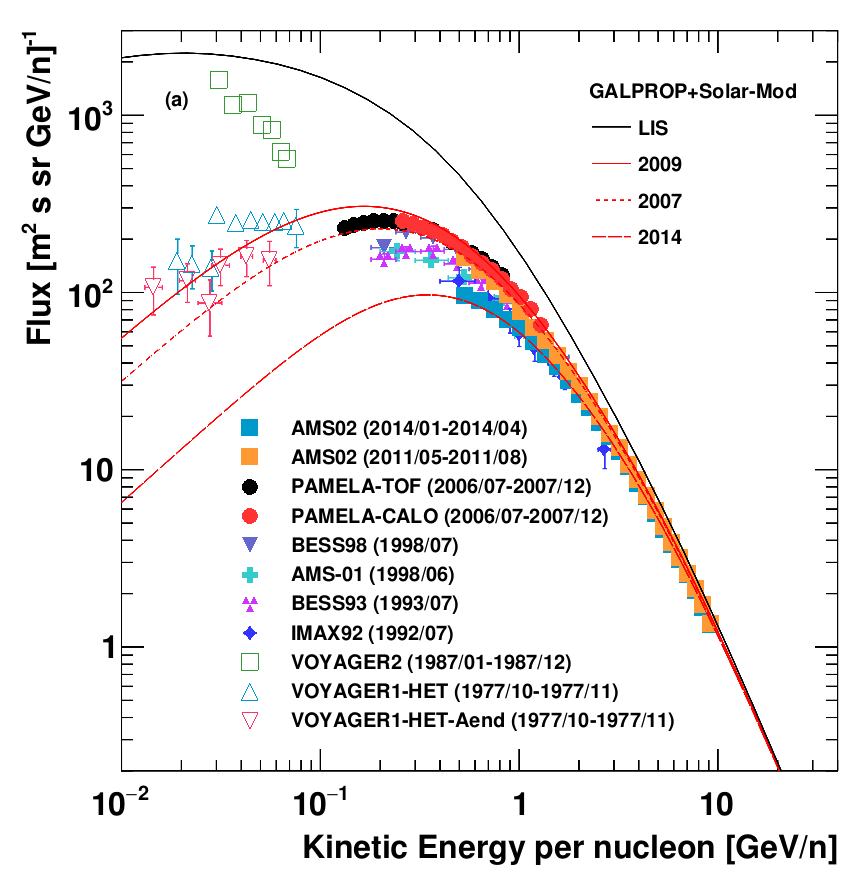}
& \includegraphics[width=0.47\linewidth]{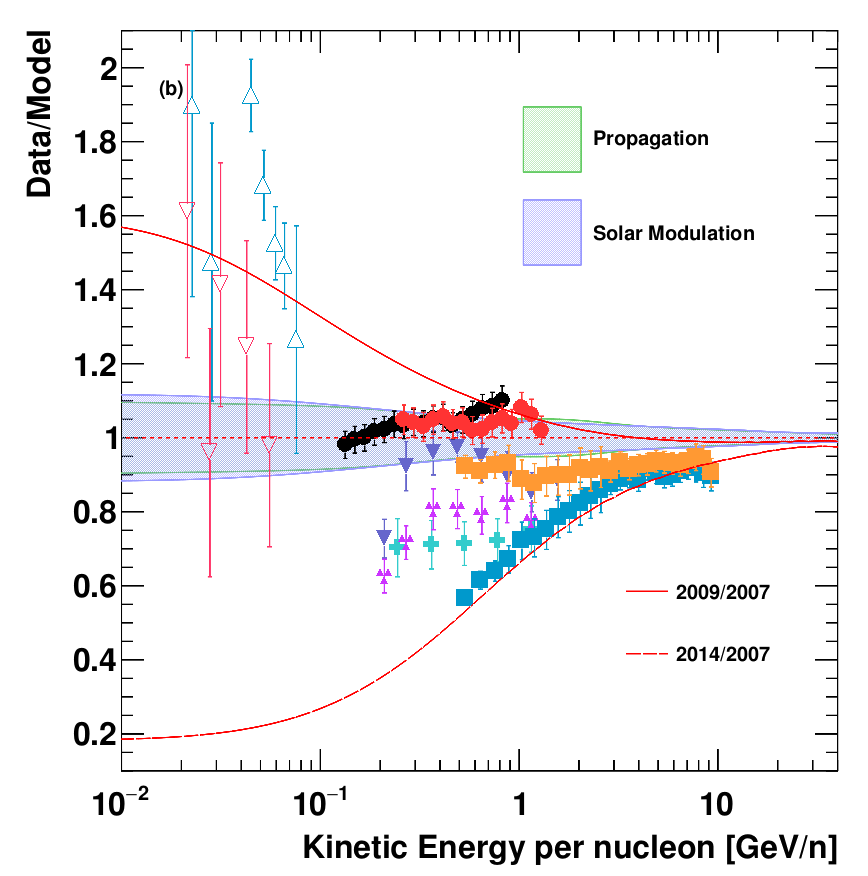}
\end{tabular}
\caption{$^{4}$He results: Plot\,(a) shows flux measurements and simulation results using {\tt GALPROP} for the LIS and three solar periods. Plot\,(b) shows the data-over-model ratio, including uncertainties from propagation, and solar modulation.}
\label{a2:fig1}
\end{figure*}

In Fig.\,\ref{a2:fig1} $^4$He LIS result (solid black line) along with modulated fluxes for the same three periods used in deuteron analysis (red lines) are compared to AMS-02\,\cite{AMSHeIsotopes}, PAMELA\,\cite{PAMELA}, BESS\,\cite{BESS93, BESS939495, BESS97, BESS98, BESS00}, AMS-01\,\cite{AMS01}, IMAX\,\cite{IMAX}, and Voyager\,\cite{Voyager1_77, Voyager2} data. As can be observed in the figure, the simulation for the period 2007 corresponding to the time when PAMELA data were measured, agrees satisfactorily with these measurements. Furthermore, the simulation for the 2014 period also agrees well with AMS-02 data for that period and converges to measurements in other periods when solar effects decrease. This level of description is expected since the propagation parameters and the solar modulation model were tuned to He data from these two experiments.

\newpage

\section*{References}
\bibliography{bibliography}
\end{document}